\documentclass[final,3p,times]{elsarticle}

\usepackage{amssymb}
\usepackage{amsmath}
\usepackage{natbib}
\usepackage{subcaption}
\usepackage{hyperref}
\usepackage{cleveref}
\hypersetup{
    colorlinks,
    linkcolor={red},
    citecolor={blue},
    urlcolor={blue},
    urlbordercolor=1 0 0
}

\begin{document}

\begin{frontmatter}



\title{Open Quantum Systems Approaches for Heavy-Ion Collisions}


\author{Alexander Rothkopf} 
\ead{akrothkopf@korea.ac.kr}

\affiliation{organization={Department of Physics, Korea University},
            city={Seoul},
            postcode={02841}, 
            country={Republic of Korea}}

\begin{abstract}
In this contribution to Strangeness in Quark Matter 2026, I review the open quantum systems approach, a modern theoretical framework for addressing the interaction of a quantum system with its environment, and highlight recent progress in its application to the understanding of in-medium heavy quarkonium in relativistic heavy-ion collisions.
\end{abstract}


\end{frontmatter}



\section{Motivation}

Understanding the interaction of a quantum system with its environment is key to shedding light on central phenomena observed in relativistic heavy-ion collisions. The quenching of highly energetic jets or the suppression of heavy quarkonia inside the quark-gluon plasma are both manifestations of the effects of a strongly coupled environment on a spatially localized probe.

Both jets and heavy quarkonia are considered ideal probes of the hot state of nuclear matter, produced in the center of a relativistic heavy-ion collision. In order to fulfill this role, they must interact with their environment such that information about the various scales present in the medium is imprinted on their internal state. Through measurement in the detectors surrounding the collision, we interrogate the probe on its internal state and infer, secondhand, the properties of the environment. Let me emphasize the reciprocity inherent in this process: the probe measures the environment while, at the same time, the environment measures the probe.

\begin{figure}
\centering
\includegraphics[scale=0.4]{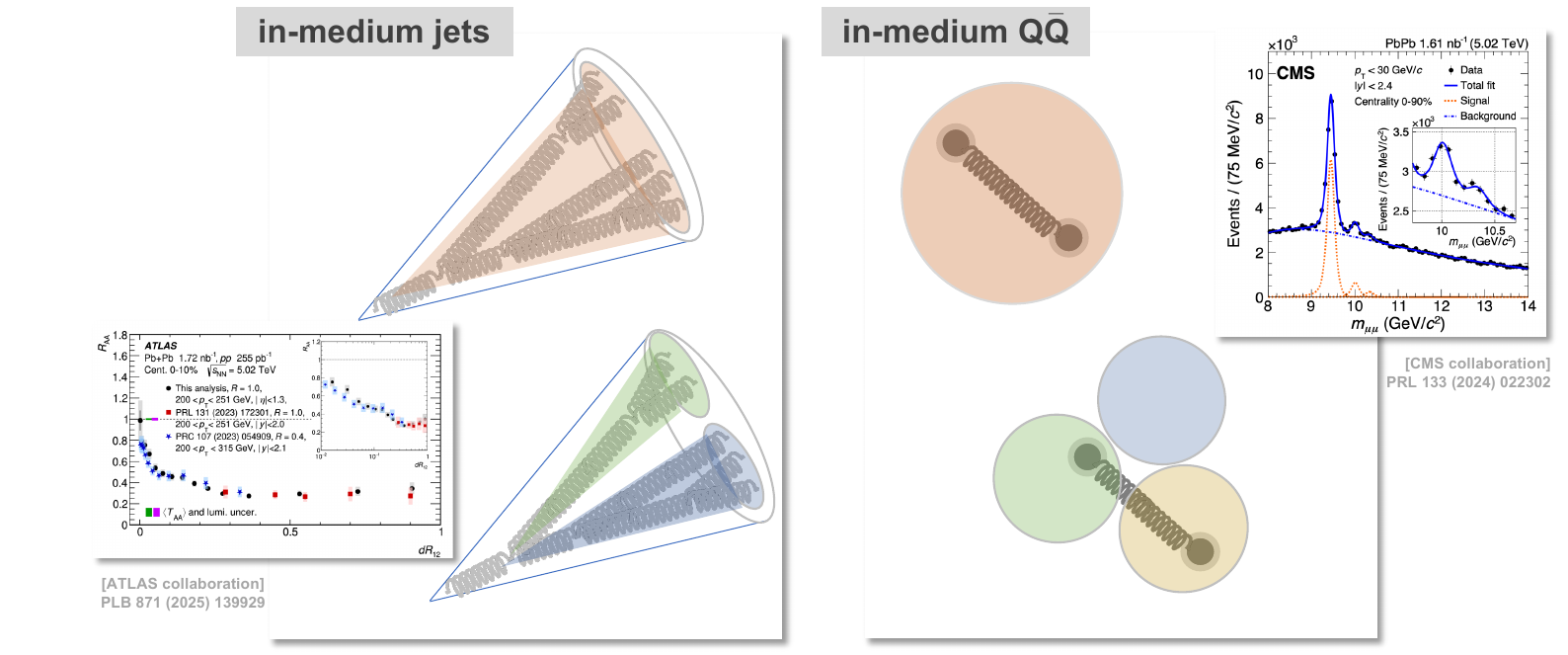}
\caption{Interaction of a jet (left) and heavy quarkonium (right) with a hot medium. Depending on the color correlation length of the environment, the substructure of the probes remains unresolved (top) or is resolved (bottom). A more highly resolving medium is able to quench jets and dissolve heavy quarkonium more efficiently as evidenced by measurements of jet $R_{AA}$ \cite{ATLAS:2025svn} (left inset) and quarkonium yields (right inset) \cite{CMS:2023lfu}.}
\label{fig:probesHIC}
\end{figure}

How does being measured affect the probe? Both jets and heavy quarkonium provide examples of the phenomenological relevance of this process. On the left-hand side of \cref{fig:probesHIC} we consider a jet propagating in a hot medium. If the color correlation length is large compared to the extent of the jet, its substructure is not resolved and the jet appears to the medium as a coherent color source. On the other hand, if the medium is more highly resolving, so that it becomes aware of the jet substructure, the jet appears as a collection of incoherent color sources that are able to shed energy more efficiently. The resulting difference in the quenching of the jet is experimentally captured by the jet nuclear modification factor shown in the left inset of \cref{fig:probesHIC} \cite{ATLAS:2025svn}. The x-axis tells us about the angular separation of the two leading subjets within the reconstructed jet, a measure for how well its substructure is resolved by the medium. The better the resolution, the more pronounced the suppression of the jet. For recent theoretical studies, see Refs.~\cite{Mehtar-Tani:2024smp,Mehtar-Tani:2025xxd,Vaidya:2026yfa}.

In the case of the bound states of a heavy quark and an antiquark a similar resolution-dependent mechanism is present, referred to as color decoherence (see e.g. discussion in \cite{Kajimoto:2017rel}). When a heavy quarkonium particle is immersed in a medium that does not resolve its substructure, interactions with the environment occur in sync for the quark and antiquark constituent, color rotating the bound singlet state into another color singlet state. However, if the color correlation length drops below the spatial extent of the $Q\bar{Q}$ system, color rotations act individually on each constituent, easily rotating a singlet into one of the various possible octet configurations. Since an octet is not attractively bound, the heavy quarkonium state dissociates. This mechanism may become active independent of the phenomenon of screening, where the bound state dissolves because the force carrier is unable to travel between the constituents. The study of quarkonium as an open quantum system has highlighted the importance of dissociation due to color decoherence, adding a new facet to our understanding of the suppression patterns observed experimentally, e.g., by the CMS collaboration \cite{CMS:2023lfu} shown in the inset on the right (see also P. Petreczky's talk \cite{Petreczky:2026sqm} on quarkonia in medium at this conference). A more detailed discussion can be found in recent theory reviews \cite{Rothkopf:2019ipj,Sharma:2021vvu,Yao:2021lus,AKAMATSU2022103932}.

The reason for the open quantum systems (OQS) approach to be ideally suited to improve our understanding of these processes is that it demystifies the measurement process itself (for a textbook see e.g. \cite{oqsbook}). It understands it as merely a manifestation of the coupling between our quantum system under scrutiny to its environment. In turn we are led to the understanding that a measurement is an inherently dynamical process, which is why the OQS framework focuses on real-time evolution. One often hears OQS mentioned in the same context as effective field theories and they do share a close relation. If there is a separation of scales between the environment and the probe system, then a simplification of the description ensues. While EFTs rely on a separation of energy scales, OQS consider the separation of time scales.

\begin{figure}
\centering
\includegraphics[scale=0.45]{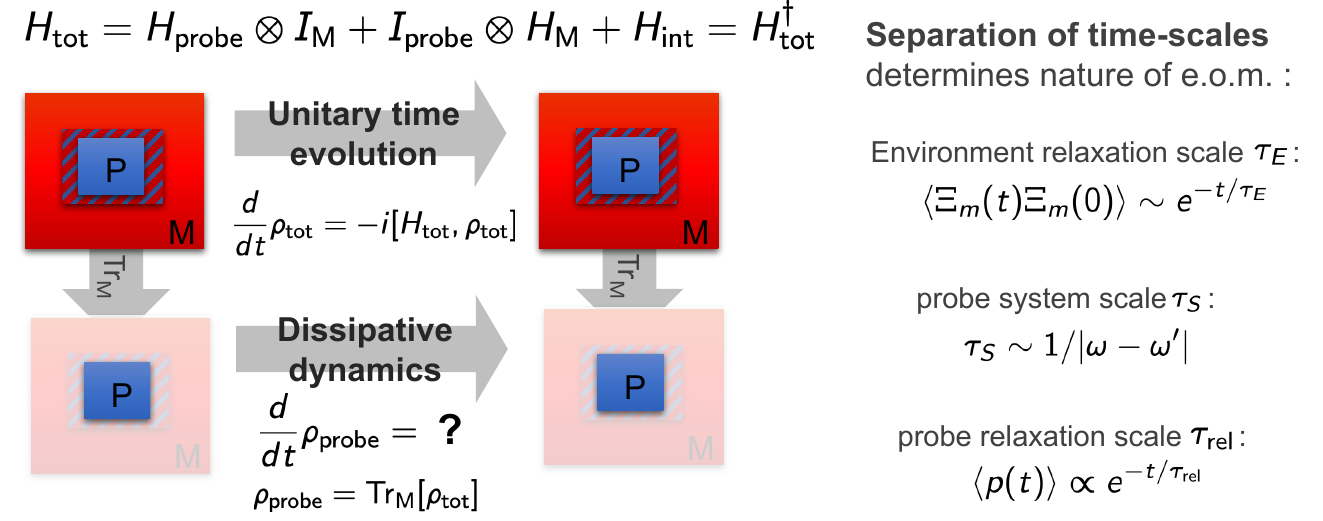}
\caption{Schematic overview of the open quantum systems framework. Its goal is to establish a master equation for the reduced density matrix $\rho_{\rm probe}$ describing dissipative evolution. Three characteristic time scales influence the form of the master equation.}
\label{fig:OQSoverview}
\end{figure}

Let me briefly describe the ingredients of the OQS framework. The conventional starting point is to consider both the probe system and the environment together as a closed system, described by a hermitian Hamiltonian $\hat H_{\rm tot}=\hat H_{\rm probe}+\hat H_{\rm env}+\hat H_{\rm int}=\hat H_{\rm tot}^{\dagger}$ and which evolves unitarily according to the von Neumann equation for the total density matrix $\partial_t \hat \rho_{\rm tot}=i[\hat \rho_{\rm tot},\hat H]$. Now if we wish to focus on the dynamics of the probe, we may formally integrate out the environment degrees of freedom, arriving at the reduced density matrix $\hat \rho_{\rm probe}={\rm Tr}_{\rm M}[\hat \rho_{\rm tot}]$. The central task for theory is to derive the evolution (master) equation of the reduced density matrix, which generally encodes dissipative dynamics, as the probe can exchange energy and momentum with the environment. As indicated in \cref{fig:OQSoverview}, the same result must be obtained from the dissipative master equation and by tracing out of the final total system density matrix.

There are three important timescales that determine the nature of the master equation. The environment relaxation scale $\tau_{\rm E}$ denotes how quickly perturbations in the environment decay. The probe intrinsic scale $\tau_{\rm S}$ describes how quickly information is updated within the probe and the so-called probe relaxation scale $\tau_{\rm rel}$ quantifies how quickly information travels between the medium and the probe. Intuitively it is the timescale after which the initial directed momentum of a point particle is randomized when exposed to the environment.

In the case where the environment updates much faster than information travels to and from the probe $(\tau_{\rm E}\ll \tau_{\rm rel})$, we may neglect memory effects, the so-called Markovian limit. In that case it was shown that any OQS can be described by a so-called Lindblad equation \cite{Lindblad:1975ef,Gorini:1975nb}
\begin{align}
\frac{d}{dt}\hat \rho_{\rm probe} = -i[\hat H_{\rm probe}^{\rm red},\hat \rho_{\rm probe}]+\sum_k \gamma_k\Big(\hat L_k \hat \rho_{\rm probe}\hat L_k^\dagger -\frac{1}{2} \big\{ \hat L_k \hat L_k^\dagger, \hat \rho_{\rm probe}\big\} \Big).
\end{align}
The operators $\hat L_k$ are called Lindblad operators and they are associated with a positive relaxation rate $\gamma_k$. Note that $\hat H_{\rm probe}^{\rm red}$ does not necessarily equal $\hat H_{\rm probe}$ as the environment may have introduced modifications of e.g. the potential acting within the probe system. While a Lindblad equation in a particular basis represents a deterministic partial differential equation for the density matrix elements, it can be rewritten (unravelled) in terms of a stochastic evolution of an ensemble of wavefunctions, in what is called quantum state diffusion \cite{Gisin:1992rxp}. This duality is similar to describing the physics of classical diffusion either via a Fokker-Planck equation or the Langevin equation. 

If one achieves to express the dynamics of a Markovian OQS in this form, it is guaranteed that the positivity of the reduced density matrix, its hermiticity and its trace are exactly conserved. 

Various Lindblad equations have been derived for systems with characteristic separation of scales. The quantum Brownian motion limit describes a system where the probe relaxation time is much larger than the environment relaxation time $\tau_{\rm R}\gg\tau_{\rm E}$ as well as the probe's intrinsic time scale $\tau_{\rm S}\gg\tau_{\rm E}$. Another relevant scenario, the quantum optical limit, emerges if the relaxation scale is large compared to the environment scale $\tau_{\rm R}\gg\tau_{\rm E}$ but the intrinsic dynamics are faster than the relaxation scale $\tau_{\rm S}\ll\tau_{\rm R}$ (see e.g. page 170 of Ref.~\cite{oqsbook}). 

\section{Quarkonium OQS ecosystem in heavy-ion collisions}
\label{sec:QQbarOQS} 

In order to understand the properties and limitations of the various OQS approaches to quarkonium, let us briefly revisit the effective field theory view of these heavy quark bound states (for a review of the $T=0$ theory see e.g. \cite{Brambilla:2004jw}). We will find that the inherent separation of energy scales (EFT) will be mirrored in timescales (OQS). Traditionally one exploits the separation between the heavy quark rest mass and two characteristic scales of strongly interacting systems, i.e. the scale of quantum fluctuations $\Lambda_{\rm QCD}/m_Q\ll1$ and temperature $T/m_Q\ll1$ (for a more detailed exposition see \cite{Ghiglieri:2012iaw,Rothkopf:2019ipj}). This allows us to treat heavy quarks non-relativistically, moving from a description in terms of four-component Dirac spinors to a theory of two sets of Weyl spinors, called Non-relativistic QCD (NRQCD). We say that modes on the hard scale $E\sim m_Q$ have been integrated out in the form of the Wilson coefficients of NRQCD. As a next step one may consider a multipole expansion to go over to a description of quarkonium in terms of color singlet and octet wavefunctions. This theory is known as potential NRQCD, since its Wilson coefficients include spatially non-local but time-independent potential terms governing the evolution of the wave functions in their respective color sectors (amended by operators which allow transitions among these sectors which do not have the form of a potential in general).

Now we are in a position to explore the OQS ecosystem for heavy quarkonium. Traditionally in-medium bottomonium was predominantly described by a deterministic Schr\"odinger equation and charmonium via rate equations (see e.g. \cite{Andronic:2015wma}). The OQS approach has made possible to construct a chain of well controlled approximations that connect these models with fundamental theory, establishing more accurate master equations on the way.

One branch of the OQS ecosystem starts from the multipole expansion underlying pNRQCD. When combined with a weak coupling approach to the environment, one can derive a Lindblad equation, which, upon additionally assuming the quantum optical limit reduces to a Boltzmann equation \cite{Yao:2018nmy,Yao:2020xzw}. If further timescale separation and proximity to equilibrium are assumed, the rate equation follows.

If one focusses on very tightly bound quarkonium states in the small dipole approximation then the weak coupling assumption of the medium can be avoided and a Lindblad equation emerges in which the dynamics is governed by a set of non-perturbative transport coefficients \cite{Brambilla:2019tpt,Brambilla:2021wkt,Brambilla:2023hkw}. Extracting these transport coefficients from lattice QCD and the relaxation of the small dipole approximation are current research directions pursued by the community (see e.g. \cite{Brambilla:2025cqy}).

One may also start from the effective description NRQCD, where quarks and antiquarks are still considered separate entities. Limiting to a weakly coupled environment at high temperatures one finds a time scale separation compatible (mostly) with the quantum Brownian motion limit \cite{AKAMATSU2022103932}, which allows one to derive a corresponding Lindblad equation \cite{Akamatsu:2014qsa}. After performing a color singlet projection and unravelling the dynamics via quantum state diffusion one finds that the quarkonium particle obeys a stochastic non-linear Schr\"odinger equation \cite{Miura:2019ssi,Miura:2022arv}, lending theory support to prior phenomenological work based on the Schr\"odinger-Langevin approach \cite{Katz:2015qja}. In the dissipationless limit the dynamics reduce to a linear stochastic Schr\"odinger equation \cite{Akamatsu:2011se} which under the adiabatic approximation yields the deterministic Schr\"odinger equation.

There exists one more branch of research which starts from a Lindblad equation derived in the context of NRQCD and which asks the question of the range of validity of the semi-classical approximation \cite{Blaizot:2017ypk,Blaizot:2018oev}. Fast decoherence would allow for a treatment with reduced numerical cost, opening the way for capturing multiple quarkonium pairs.


Let me highlight two areas of recent progress in the OQS treatment of heavy quarkonium. The first concerns the extension of the range of validity of the OQS approach. Traditionally OQS approaches covered only a limited range of relevant scales with the two regimes (quantum optical vs. quantum Brownian) sharing a limited window of overlap. Drawing on the experience of the nuclear theory community with field theory and functional approaches, the SUBATECH group has presented in Ref.~\cite{Hammou:2026bpn} work on a universal Lindblad equation which aims at covering both of the two regimes, based on an interesting factorization property of Green's functions. The TU Munich group, on the other hand, in Ref.~\cite{Brambilla:2025sis} has extended the non-perturbative Lindblad equation beyond the leading order in the multipole expansion underlying the pNRQCD framework. This now allows their master equation to reach a steady state between the quarkonium probe and the environment.

The second concerns the study of the semi-classical limit. It is known from fully quantum master equations that retain the full color structure of the quarkonium system that an extension to multiple quarkonium states suffers from a curse of dimensionality, which is absent in the semi-classical limit. Recent results focussing on uncertainty quantification for the semiclassical approximation presented in Ref.~\cite{Daddi-Hammou:2025hdz} showed promising results with reliable reproduction of the survival probability of the lowest-lying quarkonium states.

\section{OQS for discovery in heavy-ion collisions -- optimal observables}

In the third part of this contribution I would like to highlight future opportunities related to the OQS treatment of heavy quarkonium in heavy-ion collisions. We have seen the wealth of methodological developments on the theory side, when it comes to master equations for heavy quarkonium. Now it is the right moment in time to translate these developments into physics opportunities. Let me list three key challenges the community encounters:
\begin{itemize}
\item \textit{Scarcity of existing observables}: For bottomonium we have high-quality data on $R_{AA}$ but only for three states $\Upsilon(1S)$, $\Upsilon(2S)$ and recently $\Upsilon(3S)$. Compared to the observational access experiment has on probes in AMO physics this is a very small number and we must answer the question of how to optimally exploit the existing data.
\item \textit{QGP properties are not direct observables}: We only have access to the QGP via the proxy of the probe system and we must answer the question of which property (observable) of the probe is most sensitive to the environment property of interest. In other words, how to achieve an optimal thermometer, viscometer, or diffusion meter for the QGP?
\item \textit{Limited resources}: With an abundance of high-quality data from LHC and RHIC, we must decide which observables are most promising to invest human resources into. OQS can help us to decide which observable provides the best ROI, i.e. we can answer the question which observable provides the largest increase in information about the environment.
\end{itemize}

A fruitful analogy to build upon in my opinion is that of impurity physics. Whether we are interested in a heavy quarkonium state traversing the quark-gluon plasma or we are observing the Brownian motion of a heavy impurity (often called a bosonic or fermionic polaron) in an ultracold quantum gas, we utilize a probe to determine the properties of an environment that is otherwise difficult to access.

And even if the temperature scales are vastly different, the deployed mechanisms share common elements. Take e.g. the color decoherence \cite{Kajimoto:2017rel} of heavy quarkonium at $T=10^{12}\,\mathrm{K}$ as a measure of the temperature of the QGP. In a cold quantum gas setting $T=10^{-9}\,\mathrm{K}$ one might instead consider a qubit coupled to its environment and inspect the decay of spin-related observables to extract information about the system temperature as proposed in Ref.~\cite{mitchison2020situ}.

The systematic study and design of the measurement process is the purview of quantum metrology (for a recent review see \cite{montenegro2025quantum}). A key target is to estimate the error made in the measurement of an environmental property, e.g. temperature $T$ \cite{mehboudi2019thermometry} when choosing as observable the operator $\hat O$
\begin{align}
\delta T[\hat O] = \frac{ \sqrt{ \langle\hat O^2\rangle-\langle\hat O\rangle^2}}{\sqrt{N\chi^2_T[\hat O]}} , \quad \chi_T[\hat O]=\partial_\xi \left. {\rm Tr} [\hat\rho_{\rm probe}(\xi)\hat O] \right|_{\xi=T}.
\end{align}
Besides the intrinsic quantum spread of the observable in the numerator, we find in the denominator the classical dependence on the number of measurements $N$ and the operator sensitivity $\chi_T[\hat O]$. If we were to choose an operator which only weakly depends on the environmental parameter of interest, we will not be able to obtain a robust estimate.

The above relation offers a systematic approach to designing optimal observables, by expressing it as an optimization problem in the space of observables. And indeed it can be shown that optimal sensitivity is achieved via a unique quantity, the so-called symmetric logarithmic derivative (SLD), often denoted $\hat \Lambda_T$, defined via the following operator relation
\begin{align}
\Big\{ \hat \Lambda_T , \hat \rho_{\rm probe}\Big\} = 2 \partial _T \hat \rho_{\rm probe}\label{eq:SLD}.
\end{align}
The SLD is intimately connected to another key quantity of quantum metrology, the Fisher information ${\cal F}_\theta$. Indeed, it can be expressed by the variance of the SLD ${\cal F}_\theta = {\rm Var}\big[ \hat\Lambda_\theta \big]$ and intuitively speaking encodes how much we can learn about the environmental parameter $\theta$.

In practice the true SLD is often not accessible, as in general it may consist of a linear combination of infinitely many observables. However, we may ask what the optimal observable is, given a set of experimentally accessible observables using \cref{eq:SLD}.

One key limitation to explicitly constructing the SLD in the past was the need for analytical access to the reduced density matrix, in order to be able to compute the RHS $\partial _T \hat \rho_{\rm probe}$ of \cref{eq:SLD}. This severely limits the applicability of the approach to systems in nuclear physics, where at best a numerical simulation of the environment is available. 

\begin{figure}
\centering
\includegraphics[scale=0.45]{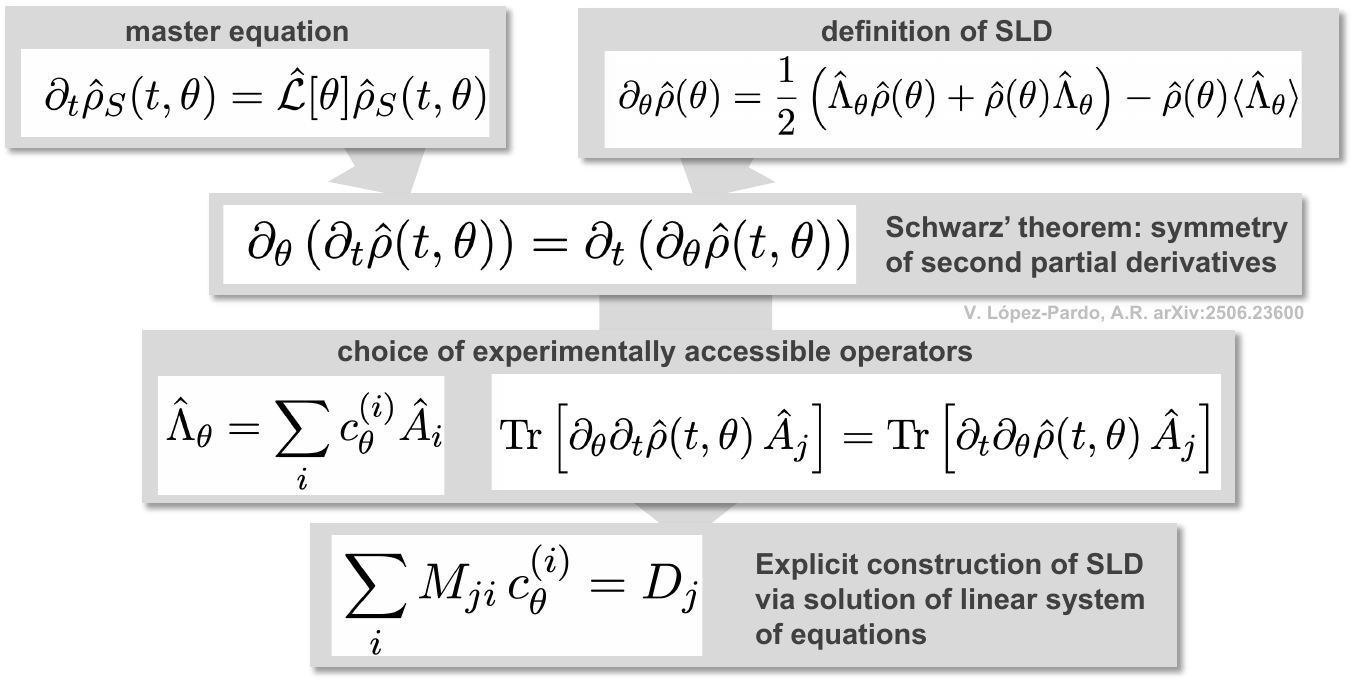}
\caption{Outline of the strategy developed in Ref.~\cite{Lopez-Pardo:2025eqe} to construct optimal observables from the master equation, encoding state-dependent information in the expectation values of experimentally accessible observables. Our approach is exact in equilibrium and approximates the true SLD in the quasi stationary approximation out of equilibrium.}
\label{fig:ConstrSLD}
\end{figure}

Addressing this methodological gap, we have made recent progress in Ref.~\cite{Lopez-Pardo:2025eqe} in devising a strategy that allows us to determine the SLD in equilibrium only from the knowledge of the generators of the master equation. That is, we do not require access to the solution of the master equation itself. All the information about the state is instead encoded in expectation values of experimentally accessible observables. Out of equilibrium our approach finds an operator which is the optimal observable in the quasistationary approximation, i.e. it amounts to the SLD if the evolution of the environment is slow compared to the relaxation scale. Technical details of the construction can be found in a recent preprint with V{\'\i}ctor L{\'o}pez-Pardo; a sketch of the procedure is given in \cref{fig:ConstrSLD}.

One key ingredient to our construction is the observation that both the master equation and the definition of the SLD are formulated in terms of partial derivatives of the density matrix. In the former, we have the time derivative; in the latter, the change w.r.t. the environmental parameter. If we apply the other derivative to each of them, then Schwarz's theorem on the symmetry of mixed partial derivatives allows us to set the two expressions equal. This establishes the connection between the master equation and the SLD \cite{Lopez-Pardo:2025eqe}. If we choose a basis of experimentally accessible operators $\hat A_i$ and express our optimal observable as a linear combination thereof $\hat \Lambda=\sum_i c^{(i)} \hat A_i$, we can convert the operator-valued equations into expressions involving only the expectation values of these accessible observables. The outcome is a linear system of equations for the expansion coefficients $c_i$ formulated solely in terms of accessible observables and the parameters of the master equation, which we can solve.

Let me demonstrate the procedure using an iconic OQS model for quantum Brownian motion, the Caldeira--Leggett model \cite{caldeira1983path}. It describes the evolution of a heavy point particle in a bath of light harmonic oscillators. Its master equation for the point particle density matrix $\hat \rho$ reads
\begin{align}
\frac{d}{dt}\hat \rho = -i [\hat H,\hat \rho] -2m\gamma T[\hat x,[\hat x,\hat \rho]] -i\gamma[\hat x,\{\hat p,\hat \rho\}]\label{eq:CLme}
\end{align}
featuring a momentum independent fluctuation term and a momentum dependent dissipation term. The properties of the
environment are encoded in this master equation through the temperature parameter $T$ and the relaxation rate $\gamma$. Our goal is to construct an optimal observable to determine these two properties.

As mentioned above, we restrict ourselves to constructing an optimal observable from within a set of experimentally accessible operators, here $\hat A_i\in\big\{ \hat x,\hat p,\hat x^2,\hat p^2,\{\hat x,\hat p\}\big\}$. Combining the information contained in the master equation \cref{eq:CLme} and the definition of the SLD \cref{eq:SLD} we solve for the optimal linear combination of the $\hat A_i$'s to determine the environment temperature $c^{(i)}_T$ and relaxation rate $c^{(i)}_\gamma$.

\begin{figure}
\centering
\includegraphics[scale=0.7]{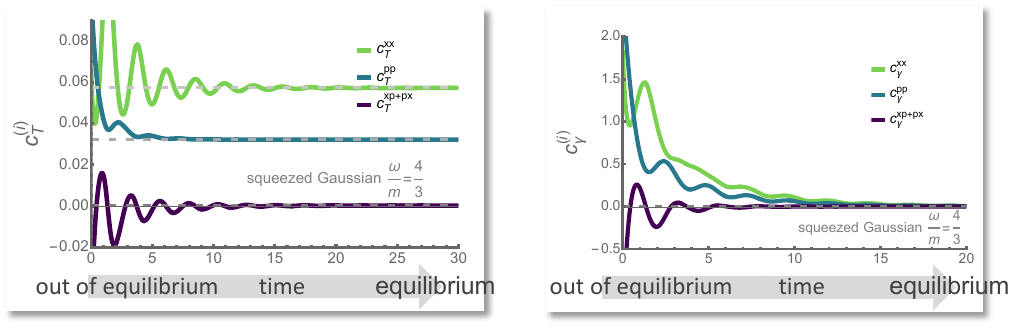}
\caption{Values obtained in the Caldeira--Leggett model for the expansion parameters $c^{(i)}$ of the optimal observables for temperature (left) and relaxation rate (right). See main text for details.}
\label{fig:CLcoeff}
\end{figure}

Starting from a squeezed Gaussian initial state, we can evolve the master equation and determine numerically the expectation values $\langle \hat A_i\rangle(t)$, which lead to the expansion coefficients shown in \cref{fig:CLcoeff}. For temperature measurements (left panel) we find that out of equilibrium $\{ \hat x,\hat p\}$ plays an important role, while as one approaches equilibrium only the position spread $\langle \hat x^2\rangle$ and momentum spread $\langle \hat p^2\rangle$ remain relevant. Our optimal observable recovers the known solution for the temperature SLD in equilibrium at late times (gray dashed lines). Interestingly, the relaxation rate optimal observable (right panel) behaves quite differently, as we remain sensitive to this non-equilibrium quantity only at early times and lose access to it at late times.

\begin{figure}
\centering
\includegraphics[scale=0.5]{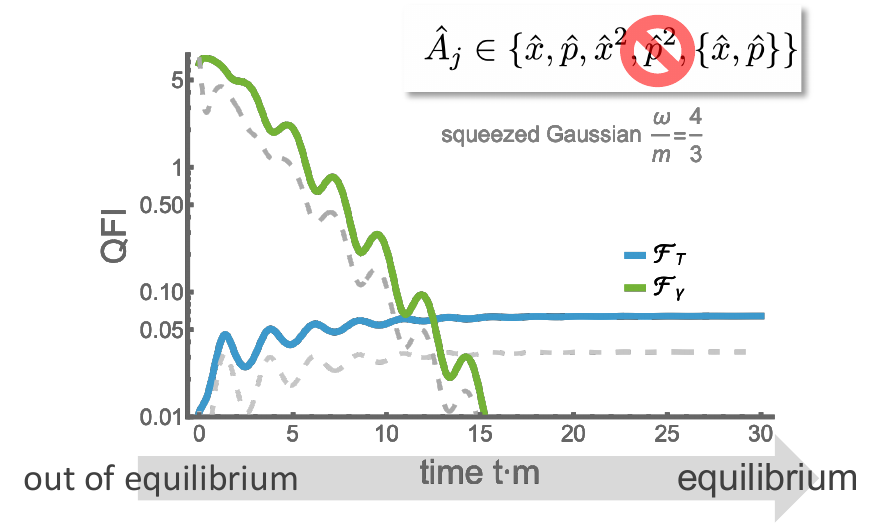}
\caption{Approximate quantum Fisher information from the variance of our optimal observables for temperature measurement (blue) and relaxation rate (green). The dashed lines denote the variance if the momentum spread observable $\hat p^2$ is removed from the set of operators contributing to the optimal observable, leading to a loss of available information.}
\label{fig:ApproxQFI}
\end{figure}

We may compute the counterpart to the Fisher information in the restricted basis of experimentally accessible operators, i.e. the variance of our optimal observable. It is an interesting quantity, as it allows us to judge which observable contributed how much information to the final result. As an example we plot the approximate QFI in \cref{fig:ApproxQFI} for both relaxation rate metrology (solid green) and thermometry (solid blue). 

The values of the accessible information behave according to our intuition. Temperature is most accurately measured if the probe is fully equilibrated with the environment, as reflected in the increase of ${\cal F}_T$ towards equilibrium. On the other hand, the relaxation rate is relevant for the approach to equilibrium. Once the system loses memory about its past, the definition of thermal equilibrium, the information about the relaxation rate that can be obtained from the probe vanishes.

While the solid lines show the values obtained by including all of the $\hat A_i$'s, the dashed lines denote the values obtained after removing the momentum spread observable $\hat p^2$. We clearly see that removing an observable lowers the amount of information we can learn about the environmental property, affecting the relaxation rate and temperature measurements differently. This crude example offers an important insight, however: by adding or removing observables, changes in the approximate QFI indicate the relative importance of each observable. Such information can be exploited to select the most promising observables among potential candidates.

All of the examples so far were obtained in a model system. The strategy to construct optimal observables, however applies verbatim to the master equations discussed in \cref{sec:QQbarOQS}.  The closest relative to Caldeira--Leggett is the pNRQCD-based master equation expressed in terms of transport coefficients. Work is ongoing to formulate the linear system to obtain the expansion coefficients for a restricted basis of observables, such as the bottomonium $\Upsilon(1S)$, $\Upsilon(2S)$ and $\Upsilon(3S)$ states. In preliminary work, however, only a qualitative assessment has been achieved, telling us that $\Upsilon(1S)$ survival appears more sensitive to the quarkonium diffusion transport coefficient, while in order to measure the transport coefficient related to in-medium potential modification, excited states must be included in the observable.

\section{Conclusion}

The open quantum systems approach constitutes a versatile framework to describe the physics of a probe coupled to an environment. It offers a modern understanding of the measurement process and provides the theory basis for quantum metrology. The OQS ecosystem for heavy quarkonium has matured over the past decade and produces steady progress in terms of improving its range of validity. 

A key aspect of quantum metrology relevant for heavy-ion collisions is the systematic construction of optimal observables, which has been formalized in the symmetric logarithmic derivative. Recent progress in the construction of an approximate SLD from the master equation supports the construction of optimal observables in more realistic systems without access to an analytic solution of the reduced density matrix. Via the approximate quantum Fisher information we may in turn gain key insights into the sensitivity of various observables to the environmental parameters of interest. The application to quarkonium master equations is work in progress.

\section*{Acknowledgements}

A.R. gladly acknowledges support by Korea University through project K2605081 \textit{Ab-initio lattice simulations of the real-time dynamics of non-relativistic fermions}, the 4th (K2024267) and 5th (K2616311) installments of the BK21 project at the Physics Department of Korea University, as well as project R252206 funded by the National Research Foundation of Korea.

\bibliographystyle{elsarticle-num}
\bibliography{OQS_HeavyIon}

\end{document}